\documentstyle[11pt,aaspp4]{article}

\received{1999 September 1}
\accepted{2000 March 27}

\slugcomment{To be appeared in AJ Vol.120, No.1 (July 2000)}

\begin{document}

\title{The Central Gas Systems of Early-Type Galaxies
Traced by Dust Feature:
Based on the HST WFPC2 Archival Images
\footnote{
Based on observations made with the NASA/ESA Hubble Space Telescope, 
obtained from the data archive at the Space Telescope Science
Institute, which is operated by the Association of Universities for
Research in Astronomy, Inc., under NASA contract NAS 5-26555.
}
}

\author{
Akihiko Tomita\altaffilmark{2},
Kentaro Aoki\altaffilmark{3,5},
Masaru Watanabe\altaffilmark{3,5,6},
Tadafumi Takata\altaffilmark{4}, and\\
Shin-ichi Ichikawa\altaffilmark{3}
}
\affil{
$^2$~
Faculty of Education, Wakayama University, Wakayama 640-8510, Japan;\\
atomita@center.wakayama-u.ac.jp\\
$^3$~
National Astronomical Observatory of Japan, Mitaka, Tokyo 181-8588, Japan;\\
aokikn@cc.nao.ac.jp,
watanabe@aquarius.plain.isas.ac.jp,
ichikawa@azuma.mtk.nao.ac.jp\\
$^4$~
Subaru Telescope, National Astronomical Observatory of Japan,
Hilo, HI 96720;\\
takata@naoj.org\\
}

\altaffiltext{5}{Present affiliation:
Japan Science and Technology Corporation,
Tokyo 102-0081, Japan.}
\altaffiltext{6}{Present address:
Institute of Space and Astronautical Science,
Sagamihara, Kanagawa 229-8510, Japan.}

\begin{abstract}
We investigated the central gas systems of E/S0 galaxies by making use
of the WFPC2 images of the Hubble Space Telescope archive.
We searched the gas systems that were traced by the dust with a new
method of making color excess images in $F555W - F814W$ ($V-I$).
Out of 25 sample galaxies, we detected gas system in 14 galaxies.
The dust was newly detected in two galaxies that were thought to
contain no dust based on single band, pre-refurbishment data.
The full extents of the gas systems are 0.1 to 3.5~kpc, and the masses
of the gas, log~$M_{\rm gas}$~[$M_{\odot}$], are 4.2 to 7.2.
The AGN activity is well correlated with existence of the gas systems.
None of galaxies without the gas systems show the AGN activity.
On the other hand, some galaxies with the gas systems show the AGN
activity;
optical AGN activities are shown in 5 out of 11 galaxies of which AGNs
are optically studied, and radio activities are shown in 6 out of 14
galaxies.
This shows that the AGN activity is driven with the gas system.
\end{abstract}


\section{Introduction}

Early-type galaxies, E/S0 galaxies, are in general gas poor.
However, some of them contain significant gas and dust in their
central regions;
it may be related to AGN activity, which make it be an interesting
object.
\cite{jaf93} (1993, 1996) presented an example;
they showed an HST image of central dust ring in NGC~4261, and they
discussed the dust ring in connection with AGN activity.
Similar dust ring structures were found in NGC~1439 and NGC~4494 by
\cite{for95} (1995) and \cite{car97} (1997), and in NGC~7052
by \cite{vdm98} (1998).

\cite{vdo95} (1995) studied statistically the central gas system which
were traced by the dust in early-type galaxies using the HST archives.
However, they used data obtained by pre-COSTAR WF/PC camera with PC
mode, and they used only one band data of $F555W$ ($V$ band) and
without color data.
Therefore, their estimates are limited by a spherical aberration of
pre-refurbishment of the HST, and a detailed morphology may be missed.
Though they pointed out a correlation between the gas systems and
the AGN activity using radio data, we thought further discussions with
post-COSTAR HST images are necessary.
Though \cite{car97} (1997) studied extensively central regions of
elliptical galaxies with $F555W-F814W$ color images by WFPC2, they did
not discuss the relation of gas systems with AGN phenomena in detail.
Recently, \cite{ver99} (1999) investigated the central dust in
early-type galaxies based on new HST observations with multibands
for a radio-loud sample.
We made use of WFPC2 archival images with two bands, $F555W$
($V$ band) and $F814W$ ($I$ band), and investigated the central gas
system traced by dust more precisely by making and analyzing color
excess images for early-type galaxies with or without AGN activities.
In section~2 we describe sample and a new method of data analysis.
We give results in section~3;
how frequent the gas systems are observed, and how much are those
masses, and what are those morphologies.
We discuss the characteristics of the gas system in connection with
AGN activity in section~4.
Conclusions are given in section~5.

\section{Data and Reduction}

\subsection{The data}

Among a large number of galaxies in the HST archive, we restricted
sample to galaxies that satisfy following criteria:
(1) objects were taken with WFPC2.
(2) objects were taken with both $F555W$ ($V$ band) and $F814W$
($I$ band) filters;
color excess image is necessary to investigate the gas system in
detail, and a combination of $F555W$ and $F814W$ gave the largest in
data frame number.
(3) position of galactic center was taken on the PC chip;
PC chip has a better spatial resolution than WF chips.
A pixel size of the PC chip was 0\farcs 0455. 
(4) objects are elliptical (E) or lenticular (S0) in morphology;
morphological type was obtained by NED (NASA Extragalactic Database).
(5) objects reside at suitable distance;
objects in the Local Group were excluded because of too much proximity
compared with other objects, and objects with heliocentric radial
velocity of $v_{\rm helio}$~$>$~3000~km~s$^{-1}$ were excluded.
The radial velocity was obtained by NED.
(6) objects with NGC or IC numbers;
we reduced number of sample in order to analyze the data in detail
intending to avoid a bias to active galaxies or peculiar galaxies.

In August 1996 we searched the HST archive and found 25 galaxies which
satisfy the criteria above.
The image data we requested and received were bias and dark
subtracted, and flat-fielded ones.
Typical exposure time of one frame was a few times 100~s.
The sample is listed in table~1;
we referred to Nearby Galaxies Catalog by \cite{tul88} (1988) for
reliable distance to galaxy and tabulated in the fifth column.
We should note that 12 of total 25 were observed under a program
of `ellipticals with kinematically distinct cores' and published by
\cite{car97} (1997);
these objects are shown by asterisks in the last column of table~1.
There is a bias for kinematically distinct cores in our sample.

\subsection{Making color image}

In processing the images, we used IRAF
\footnote[7]{IRAF is the software developed in National Optical
Astronomy Observatories.}
and STSDAS
\footnote[8]{STSDAS is the software developed in Space Telescope
Science Institute.}.
We at first removed cosmic ray events on the CCD chip by using an
STSDAS task of {\it crrej} with two to several sequential frames which
had the same pointing and pass bands.
By combining the processed frames, we obtained a set of two images,
through $F555W$ and $F814W$ filters, for each of 25 sample galaxies
with a high signal-to-noise ratio.

The extinction or color excess is due to dust in the gas system, and
we recognize the gas system by seeing the pattern of the extinction or
color excess on image.
Since gradient of intensity profile is very steep in the central
region of galaxy, it is difficult to estimate the extinction.
On the other hand, the color gradient is much flat even in the central
region of galaxy, therefore, we obtained values of the color excess
more reliably.
Thus, we made and analyzed color excess images rather than intensity
images only.
This is a new method for analyzing the gas systems.

The color image of $F555W$~$-$~$F814W$, hereafter we write it simply
as $V-I$, is derived by processing as

$$V-I\ =\ -2.5\ {\rm log}_{10}\ i(V)/i(I),$$

\noindent
where $i(V)$ and $i(I)$ are intensities, namely CCD counts, in $F555W$
and $F814W$, respectively.
The bands of $V$ and $I$ denoted here are in the ST system and they
closely correspond to the Johnson $V$ and Cousins $I$ bands,
respectively.
A detailed description is given in the WFPC2 Instrument Handbook
(\cite{bir96} 1996).
We carefully checked the position matching between images with two
bands.
We found artificial symmetric patterns in $V-I$ images of NGC~1427,
3610, and 4660 which seem to be caused by shear in position.
We shifted images by every 0.1 pixel and found the matched position,
and obtained correct $V-I$ image.

\subsection{Inspection for the gas system}

By careful inspection of the $V-I$ images, we found features caused by
dust in 14 galaxies.
The color excess image in $V-I$ was made as

$$E(V-I)\ =\ (V-I)\ -\ (V-I)_{\rm intrinsic}\ =\
-2.5\ {\rm log}\ (\ [i(V)/i(I)]\ /\ [i(V)/i(I)]_{\rm intrinsic}),$$

\noindent
where $(V-I)_{\rm intrinsic}$ means $V-I$ color without reddening due
to dust.
[$i(V)/i(I)$]$_{\rm intrinsic}$ was estimated by fitting the
$V-I$ image masking the regions where we found features by dust.
We inspected the $E(V-I)$ images if there were other features by dust
which had not been found in inspecting the $V-I$ image.
When we found new features, we again made the fitting.
Note that with circular or ring-like mask, the elliptical isophote
fitting task of {\it ellipse} in STSDAS does not fit the masked
region.
The fitting was made by an IRAF task of {\it imsurfit}.
We fitted to 1 $\times$ 1 order and 2 $\times$ 2 order curved surfaces
of spline with degree 1, spline with degree of 3, and legendre functions,
and took one which was the most successful for the fitting.
Thus, we obtained $E(V-I)$ image for each of 25 sample galaxies.

The derived $E(V-I)$ images are shown in figure~1;
gray scale is expressed in magnitude scale, and white level indicates
$E(V-I)$~$\leq$~0.0~mag, and black level indicates
$E(V-I)$~$\geq$~0.2~mag.
Size and mass of the gas system traced by the dust were inspected in
these $E(V-I)$ images.
The background noise of the $E(V-I)$ image is about 0.03~mag.
We plotted contour map with a binning of 3 $\times$ 3 pixels, which
made the contour of 0.03~mag to be three times of noise level in the
plotted figure.
We recognized the dust regions by finding places over the threshold
of 0.03~mag in the contour map with the binning.
If we found no significant patterns above this threshold, we describe
it as non-detection.
We should note that we have difficulty in detection of the dust region
in the case that way of dust distribution is the same as that of color
or intensity profile, or like a thin face-on disk.

\subsection{Deriving mass of the gas system}

We measured mass of the gas system through area-integrated $E(V-I)$
and a Galactic conversion factor.
The integrated $E(V-I)$ was measured by summation of count in $E(V-I)$
image within areas that include the dust regions we detected.
The conversion factor is obtained as follows.

\cite{vdo95} (1995) used a relation between $A_V$ and dust mass and
they took Galactic value for coefficient.
Taking gas-to-dust ratio of 100, this gives

$$M_{\rm gas}\ /\ E(B-V)\ =\ 1.4\ \times\ 10^{-2}\
[ \ {\rm g}\ {\rm cm}^{-2}\ {\rm mag}^{-1}];$$

\noindent
assuming that mass ratio of hydrogen and helium is 0.75 to 0.25, a
relation of hydrogen gas column density
($N_{\rm H~{\rm\footnotesize I}}$)
and $E(B-V)$ (e.g., \cite{dip94} 1994) gives the same result.
Taking Galactic value of $E(V-I)$~/~$E(B-V)$~=~1.6 by
\cite{rie85} (1985), we obtained

$$M_{\rm gas}\ ({\rm per\ CCD\ pixel\ on\ the\ PC\ chip})\ /\ E(V-I)\
=\ 2.0\ \times\ D^{2}\ [ \ M_{\sun}\ {\rm pix}^{-1}\ {\rm mag}^{-1}],$$

\noindent
where $D$ is distance to galaxy in the unit of Mpc.
The above derivation is based on the screen model that dust shield
lies in front of light source in the line of sight.
As a matter of fact, the gas system lies inside the galaxy.
We doubled the color excess in calculating mass of the gas system;
an error of a factor of a few times ten percent is considered to be
introduced through this procedure.
Thus, we used an equation

$$M_{\rm gas}\ [M_{\sun}]\ =\ 4.0\ \times\ D^2\
\int_{\rm area} E(V-I)\ d({\rm pixel}).$$

\noindent
Including the errors introduced by using Galactic value and in
summation of $E(V-I)$, total error in logarithmic scale in above
equation is considered to be about 0.3 to 0.5 depending on
signal-to-noise ratio and size of the region.

\section{Results}

\subsection{Frequency of the gas system}

In table~2, we summarize the detection of the gas system traced by
dust.
The gas systems were detected in 14 out of 25 sample galaxies;
more than a half, 56\%, of early-type galaxies shows the gas system in
the central region.
This result is consistent with result by \cite{vdo95} (1995) that the
gas systems were detected in 48\% of their sample galaxies.
Among our sample, all of the gas systems which were detected by
\cite{vdo95} (1995) were found by us.
Among their sample, we newly detected the dust feature in NGC~3377 and
NGC~3608 which has not been detected.
This is because we used images taken after the 1993 refurbishment
mission and a new method of inspecting color excess images.
Though we saw a possible dust feature in NGC~4458, the criterion
described in subsection~2.3 was not satisfied in this case.
We checked a distance bias in the detection rate.
We detected the dust features in 4 out of 6 galaxies with
$D \leq 10$~Mpc, 6 out of 13 galaxies with $10 < D \leq 20$~Mpc, and
4 out of 6 galaxies with $D > 20$~Mpc.
Though the number of sample is small and the statistics are poor, the
frequency does not show a dependence on distance.

For galaxies with the kinematically distinct cores, the detection
rate of the gas system is 8 out of 12, and for other galaxies,
6 out of 13.
Though the number of sample is small and the statistics are poor,
it seems that the detection rate for galaxies with the kinematically
distinct cores is a little higher than or the same as that for others.
\cite{ver99} (1999) detected the dust features in 17 out of 19
radio-loud E/S0 galaxies. A high detection rate for galaxies with AGNs
is discussed in section~4.

\subsection{Size of the gas system}

The full extent of the dust distribution, $d$, was measured and is
listed in the fourth column of table~2.
The smallest size is 0.08~kpc for NGC~4494, and the largest is 3.5~kpc
for NGC~4589.
The dust distribution seems to extend to out of the frame for
NGC~4589, and possibly for NGC~3377 as well.

\subsection{Mass of the gas system}

Mass of the gas system traced by the dust, $M_{\rm gas}$, was derived
as is explained in subsection~2.4, and results are tabulated in the
fifth column of table~2.
The smallest mass is log~$M_{\rm gas}$~[$M_{\odot}$] =~4.2 for
NGC~4472 and the largest mass is 7.2 for NGC~4526.
Note that for NGC~4476 and NGC~4526, there are many 
H\,{\footnotesize II} regions on the dust, thus, the derived masses
seem to be underestimated and to contain larger errors.
The upper limit of gas mass for objects without detection is
considered to be about 3.5 in logarithmic solar mass unit.

\cite{vdo95} (1995) reported masses of the gas systems, and most of
their sample is included in our sample.
Their estimation of mass is tabulated in the sixth column of table~2;
these values are converted using distances given by
\cite{tul88} (1988) which we show in table~1.
In some galaxies, our results are consistent with results by
\cite{vdo95} (1995), and in other galaxies our results are about an
order of magnitude larger than those by \cite{vdo95} (1995).
For instance, we derived log~$M_{\rm gas}$~[$M_{\odot}$] =~5.5 for
IC~1459, while it was 4.9 by \cite{vdo95} (1995) after the conversion
considering the distance.
\cite{vdo95} (1995) described that the morphology of the gas system to
be `warped lane'.
We found extended dust distribution around the warped lane
(see the first paragraph of subsection~3.4).
Limiting to the lane region, our estimation was
log~$M_{\rm gas}$~[$M_{\odot}$] =~4.7, which is consistent with the
result by \cite{vdo95} (1995) considering the error.
Taking a new method, the mass estimation is revised.

\subsection{Morphology of the gas system}

In inspecting the dust regions as described in subsection~2.3, we
obtained the contour plots with the threshold of
$E(V-I)$ =~0.03~mag.
Six samples, NGC~1439, 3379, 4476, 4494, 4526, and 5322, have the
outer contours of $E(V-I)$ =~0.03~mag that could be fitted by an
ellipse with its center on the galactic center.
Others have irregular outer contours and barycenters of the dust
distributions are off the galactic centers.
Though the signal-to-noise ratio is poor for NGC~3608, by a careful
image inspection we found the dust region in NGC~3608 to belong to the
former group.
We call the former group `disky' and the latter group `irregular' in
morphology of the gas system.
Though we assigned `irregular' for IC~1459, it has also a `disky'
feature.
With a contour of $E(V-I)$ of 0.1~mag, we found a disky component
(see the second paragraph in subsection~3.3). 
\cite{ver99} (1999) also made a classification of the dust features.
Our disky category corresponds to their classes of disks and lanes.
We should note their argue that difference between disks and lanes
is not solely due to different viewing angles.

In figure~2, we plot the size and mass distributions of the gas
systems.
The signs of plots indicate morphologies of the gas systems;
a circle denotes the disky type and an asterisk denotes the irregular
type.
For the irregular group, the size and mass frequency are flat;
log~$M_{\rm gas}$~[$M_{\odot}$] is from 4.2 to 7.1, and $d$ is from
0.1 to 3.5~kpc.
On the other hand, for the disky group, they are bimodal;
5 out of 7 samples have log~$M_{\rm gas}$~[$M_{\odot}$] of
$5.0 \pm 0.6$ and $d$ of $0.24 \pm 0.16$~kpc, while other two samples,
NGC~4476 and NGC~4526, have log~$M_{\rm gas}$~[$M_{\odot}$] of
$7.1 \pm 0.1$ and $d$ of $2.1 \pm 0.3$~kpc.
We call the former `small disk' and the latter `large disk'.
In figure~2, we plot the small disks with open circles, and the large
disks with filled circles.
The large disks are only samples which have the
H\,{\footnotesize II} regions.
Considering also that morphologies of the host galaxies with the large
disks are S0's, the large disks seem to be normal gas disks which are
commonly seen in spiral galaxies.
The central gas ring in NGC~4261 presented by
\cite{jaf93} (1993, 1996) would belong to the small disk group;
this object is not included in our sample.
Radii of disks in our small disk group sample are from 40 to 260~pc,
which are comparable to the radius of the ring in NGC~4261 of 140~pc;
we converted a result by \cite{jaf93} (1993) using a distance of
35.1~Mpc given in \cite{tul88} (1988) instead of 14.7~Mpc which was
used in \cite{jaf93} (1993).

\subsection{Blue nuclei and nuclear disks}

In some $E(V-I)$ images, we found point-like nuclei.
\cite{car97} (1997) and \cite{ver99} (1999) also reported these
objects.
These are found in NGC~4278, 4365, 4458, 4476, 4660, and 7457, and
IC~1459.
The nucleus has a slightly bluer color than surrounding area, and the
color excesses are of order of 0.1~mag for NGC~4278, 4476, and 7457,
and IC~1459, and of order of 0.01~mag for others.
In NGC~3115, 3377, 3384, 3610, and 4621, CCD counts near the nuclei
are saturated, thus, it is not clear whether these have the blue
nuclei.
Since the blue nucleus has a cuspy intensity peak, most of the
saturated nuclei may have the blue nuclei.

By inspecting the $V$-band images, we found structures like nuclear
disks with size of 100~pc scale in NGC~3115, 4550, and 4621.
The disk-like structure has a slightly bluer color than surrounding
area, and the color excesses are of order of 0.1~mag for NGC~4550, and
of order of 0.01~mag for others.
Therefore, we recognized the structures also in $E(V-I)$ images.
The structure in NGC~3115 was previously reported by
\cite{kor96} (1996).

\section{Discussion}

\cite{vdo95} (1995) showed that the gas systems came from outside the
galaxies by presenting a misalignment of position angels for the gas
system and the host galaxy.
We discuss here the connection of the infalling gas systems with the
AGN activity.

\cite{ho97} (1997) investigated optical AGN properties of nearby
galaxies.
The AGN properties for our sample galaxies are listed in the seventh
column of table~2.
We grouped result by \cite{ho97} (1997) into three:
AGN group which contains objects with L1.9, L2 (L for LINER),
S2 (S for Seyfert), and T2 (T for Transition between LINER and 
H\,{\footnotesize II} nucleus), and
includes objects with the sign of `:', non-AGN group which contains
objects with NO and H (H for H\,{\footnotesize II} nucleus),
and objects with the sign of `::' because of very weak emission lines,
and `no data' group which contains objects without observations by
\cite{ho97} (1997), which is signed as `---' in table~2.
Among 25 sample galaxies, 20 galaxies were observed by
\cite{ho97} (1997).
The optical AGN activities are found in 1 out of 6 galaxies with
$D \leq 10$~Mpc, 1 out of 8 galaxies with $10 < D \leq 20$~Mpc, and
3 out of 6 galaxies with $D > 20$~Mpc.
There is an indication of little distance bias in the AGN detection
that it is easier to detect the AGN activities in nearer galaxies.
The existence of the AGN activity is correlated with existence of the
gas systems.
None of 9 galaxies without detection of the gas systems are grouped in
AGN group, and 5 out of 11 galaxies with gas systems shows AGN
activity.
This indicates that the gas system which can be found in the HST WFPC2
archival image is necessary for the present AGN activity.

Radio emission is considered to be related to the AGN.
The 4.85~MHz survey in the northern hemisphere was made by
\cite{gre91} (1991) using Green Bank 91m radio telescope and compiled
as 87GB catalog;
among our sample, 24 galaxies except IC~1459 are included in this
survey region.
NGC~4278, 4472, 4589, and 5322 were detected in this survey.
4.85~MHz survey in the southern hemisphere was made using Parkes 64m
radio telescope and compiled as Parkes-MIT-NRAO (PMN) catalog;
IC~1459 was detected and described in \cite{wri96} (1996).
According to NED, NGC~5813 is another radio source detected by
\cite{whi92} (1992), though it is not cataloged in 87GB catalog.
We took above six galaxies as radio galaxies and indicated in the
eighth column of table~2.
The radio activities are found in 1 out of 6 galaxies with
$D \leq 10$~Mpc, 2 out of 13 galaxies with $10 < D \leq 20$~Mpc, and
3 out of 6 galaxies with $D > 20$~Mpc.
There is an indication of little distance bias in the radio detection
that it is easier to detect the radio in nearer galaxies.
Though being a radio galaxy is not so tightly correlated with the
optical AGN properties by \cite{ho97} (1997), correlation of the
existence of the gas system with being a radio galaxy is similar to
that with the optical AGN properties.
None of 11 galaxies without the gas systems are radio galaxies and
6 out of 14 galaxies with gas systems is radio galaxy.
\cite{vdo95} (1995) gave similar results.
\cite{ver99} (1999) made observations of 19 nearby radio-loud
early-type galaxies using the HST WFPC2, and reported that they
detected the dust in all but two galaxies.
This supports the high detection rate of the gas system in galaxies
with the AGN.

The gas mass seems to be independent of the AGN existence.
In the optical activity, those with the AGN have 
log~$M_{\rm gas}$~[$M_{\odot}$] =~4.8 to 7.1 and those without the AGN
have 4.2 to 7.2;
in the radio activity, the radio galaxies have 4.2 to 7.1 and others
have 4.3 to 7.2.
Morphology of the gas system shows rather clear contrast in the AGN
existence, though the statistics are poor.
In the optical activity, within the \cite{ho97} (1997) sample, one of
four disky gas systems, and four of six irregular gas systems have the
AGNs;
in the radio activity, one of seven disky gas systems, and five of
seven irregular gas systems are radio galaxies.
The gas system with an irregular morphology may be infalling more
effectively than that with a disky morphology.
This suggests that efficiency in infalling of the gas to the nucleus
is correlated in driving the AGN activity.

The blue nucleus may be AGN or young star cluster originated from the
gas system.
However, existence of the blue nucleus is not correlated with
existence of AGN, and not correlated with existence or morphology of
the gas system either;
its nature is unclear.
Since NGC~4550 possesses a type-II LINER, the nuclear disk-like
structure in NGC~4550 may be strong light beams escaping from nucleus
to the direction which is perpendicular to line of sight considering
the blueness of the structure.
However, the same cases are not seen in other type-II LINERs.

\section{Conclusions}

We investigated the gas systems in the central regions of 25 E/S0
galaxies by analyzing the color excess images made from HST archives.
The conclusions are:

\noindent
1. 
We detect the central gas systems that can be traced by the dust in
14 out of 25 sample galaxies.
We find the gas system in two galaxies that were reported to be
dust-free by \cite{vdo95} (1995) which was based on single band,
pre-refurbishment data.

\noindent
2. The morphologies of the gas systems are grouped into three:
small disk, large disk, and irregular system.
For small disk, full extent of $d$ is 0.1 to 0.5~kpc and
gas mass of log~$M_{\rm gas}$~[$M_{\odot}$] is 4.3 to 5.9.
For large disk, $d$ is about 2~kpc and
log~$M_{\rm gas}$~[$M_{\odot}$] is about 7.
For irregular system, $d$ is 0.1 to 3.5~kpc and
log~$M_{\rm gas}$~[$M_{\odot}$] is 4.2 to 7.1.

\noindent
3. The AGN activity is well correlated with existence of the gas
system which can be found in the HST WFPC2 archival image.

\noindent
4. The AGN activity is triggered by an infalling gas system with a
probability of about 50\%, and the probability is higher if the
morphology of the gas system is seen as irregular type, though the
statistics are poor.

\acknowledgments
We thank an anonymous referee for careful reading and making valuable
comments, which improved this paper very much.
This research was financially supported by joint research of
National Astronomical Observatory of Japan in 1997 and 1998.
This research made use of Astronomical Data Analysis Center of
National Astronomical Observatory of Japan.
This research made use of the NASA/IPAC Extragalactic Database (NED)
which is operated by the Jet Propulsion Laboratory, Caltech,
under contact with the National Aeronautics and Space Administration.

\clearpage

\begin{figure}
\caption{
$E(V-I)$ images of 25 sample galaxies.
Gray Scale is expressed in magnitude scale, and white level indicates
a color excess of 0~mag or less and black level indicates a color
excess of 0.2~mag or more.
Field of view is central 256~$\times$~256 pixels, which gives
11\farcs 6~$\times$~11\farcs 6, or central
512~$\times$~512 pixels, which gives
23\farcs 3~$\times$~23\farcs 3 for objects with large-sized gas
systems.
The direction of north and east is indicated by a sign of arrow and
bar;
the direction of the arrow shows the north.
The length of the arrow corresponds to $5''$.
(a) NGC~1427,
(b) NGC~1439,
(c) NGC~3115,
(d) NGC~3377; the up-turn feature at upper edge is artificial,
(e) NGC~3379,
(f) NGC~3384,
(g) NGC~3608,
(h) NGC~3610, and
(i) NGC~4278.
}
\label{fig1-1}
\figurenum{1}
\end{figure}

\begin{figure}
\caption{
Continued.
(j) NGC~4365,
(k) NGC~4406,
(l) NGC~4458,
(m) NGC~4472,
(n) NGC~4476,
(o) NGC~4494,
(p) NGC~4526,
(q) NGC~4550, and
(r) NGC~4589.
}
\label{fig1-2}
\figurenum{1}
\end{figure}

\begin{figure}
\caption{
Continued.
(s) NGC~4621,
(t) NGC~4660,
(u) NGC~5322,
(v) NGC~5813,
(w) NGC~5982,
(x) IC~1459, and
(y) NGC~7457.
}
\label{fig1-3}
\end{figure}

\begin{figure}
\caption{
The diagram showing size and mass distributions of fourteen gas
systems detected by us.
Open circle indicates the gas system with a small disk morphology,
filled circle indicates that with a large disk morphology, and
asterisk indicates that with irregular morphology.
}
\label{fig2}
\end{figure}

\end{document}


\begin{table*}[t]
\begin{center}
Table~2. \hspace{4pt} RESULT OF MEASUREMENTS OF THE GAS SYSTEM\\
\end{center}
\vspace{6pt}
\begin{tabular*}{\textwidth}
 {@{\hspace{\tabcolsep}\extracolsep{\fill}}rcclccccl}
\tableline \tableline \\
(1)&  (2) &  (3) &  (4) &   (5)           & (6)  & (7) & (8)  & (9)      \\
No.& Name & Dust & Size &log~$M_{\rm gas}$&vD\&F &AGN? &radio?& comments \\
   &  &morphology&[kpc] & [$M_\odot$]     &      &     &      &          \\
\tableline \\
 1 & NGC~1427 & NO         &      &     &     & ---    &     &          \\
 2 & NGC~1439 & small disk & 0.31 & 5.4 & 3.6 & ---    &     &          \\
 3 & NGC~3115 & NO         &      &     &  NO & NO     &     & sat, disk\\
 4 & NGC~3377 & irregular  & 0.5  & 4.7 &  NO & NO     &     & sat      \\
 5 & NGC~3379 & small disk & 0.11 & 4.3 & 3.5 & L2/T2::&     &          \\
 6 & NGC~3384 & NO         &      &     &  NO & NO     &     & sat      \\
 7 & NGC~3608 & small disk & 0.19 & 4.8 &  NO & L2/S2: &     &          \\
 8 & NGC~3610 & NO         &      &     &     & NO     &     & sat      \\
 9 & NGC~4278 & irregular  & 1.0  & 5.9 & 5.3 & L1.9   & yes & nuc      \\
10 & NGC~4365 & NO         &      &     &  NO & NO     &     & nuc      \\
11 & NGC~4406 & NO         &      &     &     & NO     &     &          \\
12 & NGC~4458 & NO         &      &     &     & ---    &     & nuc      \\
13 & NGC~4472 & irregular  & 0.1  & 4.2 & 3.1 & S2::   & yes &          \\
14 & NGC~4476 & large disk & 1.8  & 7.0 & 6.1 & ---    &     & nuc      \\
15 & NGC~4494 & small disk & 0.08 & 4.5 & 3.5 & L2::   &     &          \\
16 & NGC~4526 & large disk & 2.4  & 7.2 &     & H      &     &          \\
17 & NGC~4550 & irregular  & 1.6  & 6.3 & 5.4 & L2     &     & disk     \\
18 & NGC~4589 & irregular  & 3.5  & 7.1 & 5.8 & L2     & yes &          \\
19 & NGC~4621 & NO         &      &     &  NO & NO     &     & sat, disk\\
20 & NGC~4660 & NO         &      &     &     & NO     &     & nuc      \\
21 & NGC~5322 & small disk & 0.52 & 5.9 & 5.7 & L2::   & yes &          \\
22 & NGC~5813 & irregular  & 1.9  & 6.6 & 4.9 & L2:    & yes &          \\
23 & NGC~5982 & NO         &      &     &  NO & L2::   &     &          \\
24 & IC~1459  & irregular  & 0.4  & 5.5 & 4.9 & ---    & yes & nuc      \\
25 & NGC~7457 & NO         &      &     &  NO & NO     &     & nuc      \\
\tableline \\
\end{tabular*}
Notes: \\
Column (4):
Full extent of the gas system.
Column (5):
Upper limit for without detection is estimated about 3.5.
Column (6):
Mass estimated by van Dokkum and Franx (1995) in logarithmic solar
mass unit, and converted using distances given by Tully (1988).
`NO' means that they did not detect the gas system, and blank
means it was not included in their sample.
Column (7):
AGN property is taken from Ho et al.\ (1997).
NO: No emissions were detected neither around the H$\alpha$ nor the
H$\beta$ lines.
L1.9: LINER with a weak broad-line component in the H$\alpha$ line.
L2: LINER without the broad-line component.
S2: Type-II Seyfert nucleus.
H: H\,{\footnotesize II} nucleus.
T2: Transition object between LINER and H\,{\footnotesize II} nucleus.
The sign of `:' means the classification is uncertain, and `::'
means the classification is highly uncertain.
Objects that are not included in Ho et al.\ (1997) are denoted as
`---'.
Column(9):
sat: CCD count near the nucleus is saturated.
nuc: A blue point-like source is seen at the nucleus of galaxy.
disk: A nuclear disk-like structure is seen.

\end{table*}